\documentclass[aps,prl,floatfix,showpacs,twocolumn,tightenlines,superscriptaddress]{revtex4}

\usepackage{amsmath}
\usepackage{graphicx}
\usepackage{epsfig}

 \newcommand{\ket}[1]{ {\left| #1 \right\rangle} }

\begin{document}

\title{A high-efficiency quantum non-demolition single photon number resolving detector}

\author{W.\ J.\ Munro}\email{bill.munro@hp.com}
\affiliation{Hewlett-Packard Laboratories, Filton Road, Stoke Gifford, Bristol BS34 8QZ, United Kingdom}
\affiliation{National Institute of Informatics, 2-1-2 Hitotsubashi, Chiyoda-ku, Tokyo 101-8430, Japan}

\author{Kae Nemoto}\email{nemoto@nii.ac.jp}
\affiliation{National Institute of Informatics, 2-1-2 Hitotsubashi, Chiyoda-ku, Tokyo 101-8430, Japan}

\author{R.\ G.\ Beausoleil}
 \affiliation{Hewlett-Packard Laboratories, 13837 175$^\textrm{th}$ Pl.\ NE, Redmond, WA 98052--2180, USA}
\author{T.\ P.\ Spiller}
\affiliation{Hewlett-Packard Laboratories, Filton Road, Stoke Gifford,  Bristol BS34 8QZ, United Kingdom}
\date{October 01, 2003}
\begin{abstract}
We discuss a novel approach to the problem of creating a
photon number resolving detector using the giant Kerr
nonlinearities available in electromagnetically induced
transparency. Our scheme can implement a photon number quantum
non-demolition measurement with high efficiency ($\sim$99\%) using
less than 1600 atoms embedded in a dielectric waveguide.
\end{abstract}


\maketitle
In recent years we have seen signs of a new technological
revolution, caused by a paradigm shift to information 
processing using the laws of quantum physics. One natural 
architecture for realising quantum information processing 
(QIP) technology would be to use states of light  as the 
information processing medium. There have been significance 
developments in all optical QIP following the recent
discovery by Knill, Laflamme and Milburn that passive linear
optics, photo-detectors, and single photon sources can be 
used to create massive reversible nonlinearities\cite{KLM}. 
Such nonlinearities are an essential requirement for 
optical QIP and many communication applications. These
nonlinearities allow efficient gate operations to be 
performed. In principle, fundamental operations such 
as the nonlinear sign shift and CNOT gates have been demonstrated
experimentally\cite{Pittman03,OBrien03,Gasparoni04}.
However, such operations are relatively inefficient 
(they have a probability of success significantly 
less than 50\%) and so are not scalable, due primarily 
to the current state of the art in single photon sources 
and detectors. Good progress is being made on the 
development of single photon sources\cite{yamamoto1,yamamoto2}. 
Absorptive single photon resolution detection
is possible\cite{miller,rehacek,waks03},
with efficiencies up to $\sim$90\% (visible spectrum)
and $\sim$30\% (infrared, microwaves). However, {\it true} 
universal optical QIP will require significant further
improvements in detector efficiencies, which will likely
require a drastic change of approach to detection
technology\cite{waks03,james02,iammoglu02}.

In this letter, we propose an implementation of the quantum
non-demolition (QND) single-photon detection scheme originally
described by Imoto, Haus and Yamamoto\cite{imot85}, with the
required optical nonlinearity provided by the giant Kerr effect
achievable with AC Stark-shifted electromagnetically induced
transparency (EIT)\cite{Imamoglu96}. We show below that the scheme
uses about 1600 EIT atoms and a weak pulse in the probe mode to
achieve an error probability less than 1\%. The effect of the QND
measurement in turn means that signal photons are not destroyed
and can be reused if required\cite{grang98}. Furthermore, for a
signal mode in a superposition state (such as a weak coherent
state) the number-resolving QND measurement projects the signal
mode into a definite number state\cite{Milburn 1984}, so the
detector can be used as a heralded source of number eigenstates.
We focus on EIT as an example here since it has already been used
successfully to demonstrate cross-Kerr nonlinearities at low
light levels in rubidium\cite{harr99,harris04}. From the perspective
of realising our detector application, we concentrate
on potential condensed-matter mechanisms of EIT.
However, clearly any system capable of producing a comparable form and
strength of Kerr interaction can be used, including optical
fibers\cite{li04}, silica whispering-gallery
micro-resonators\cite{kipp04} and cavity QED
systems\cite{grang98,kimble95}. 

Before we begin our detailed discussion of the EIT detection
scheme, we first consider the photon number QND measurement using
a cross-Kerr nonlinearity\cite{imot85,howe00}. The Kerr
Hamiltonian has the canonical form
\begin{equation}\label{crossKerreq}
H_\text{QND}= \hbar \chi a^\dagger a c^\dagger c ,
\end{equation}
where the signal (probe) mode has the creation and destruction
operators given by $a^\dagger, a$ ($c^\dagger, c$) respectively,
and $\chi$ is the strength of the nonlinearity. If the signal
field contains $n_a$ photons and the probe field is in an initial
coherent state with amplitude $\alpha_c$, the cross-Kerr optical
nonlinearity causes the combined system to evolve as
\begin{equation}\label{phase-shift}
|\Psi(t)\rangle_{out} = e^{i \chi t a^\dagger a c^\dagger c}
|n_a\rangle|\alpha_c\rangle =|n_a\rangle|\alpha_c e^{i n_a \chi
t}\rangle .
\end{equation}
We observe immediately that the Fock state $|n_a\rangle$ is
unaffected by the interaction, but the coherent state
$|\alpha_c\rangle$ picks up a phase shift directly proportional to
the number of photons $n_a$ in the $|n_a\rangle$ state. If we
measure this phase shift using a homodyne measurement (depicted
\begin{figure}[!htb]
\includegraphics[scale=0.45]{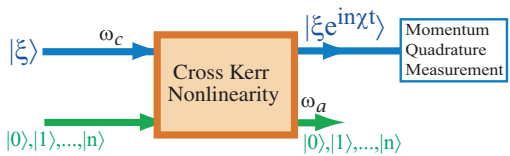}
\caption{Schematic diagram of a photon number quantum
non-demolition detector based on a cross-Kerr optical
nonlinearity\cite{imot85}. The two inputs are a Fock state
$|n_a\rangle$ (with $n_a=0,1,$..) in the signal mode $a$ and a
coherent state with real amplitude $\alpha_c$ in the probe mode
$c$. The presence of photons in mode $a$ causes a phase shift on
the coherent state $|\alpha_c\rangle$ directly proportional to
$n_a$ which can be determined with a momentum quadrature
measurement.}
\label{qnd_ideal}
\end{figure}
schematically in Fig.~\ref{qnd_ideal},
we can infer the number of photons in the signal mode $a$. The
homodyne apparatus allows measurement of the quadrature operator
$\hat{x}(\phi) \equiv c e^{i \phi} +c^\dagger e^{-i \phi}$, with
an expected result
$\langle \hat{x}(\phi)\rangle = 2 \text{Re}\left[\alpha_c\right]
\cos \delta +i 2 \text{Im}\left[\alpha_c \right]\sin \delta$,
where $\delta=\phi + n_a \chi t$. For a real initial $\alpha_c$, a
highly efficient homodyne measurement of the momentum quadrature
$Y \equiv \hat{x}(\pi/2)$ would yield the signal $\langle Y
\rangle = 2 \alpha_c \sin \left(n_a \chi t\right)$ with a variance
of one, thus giving a signal-to-noise ratio of $\text{SNR}_Y = 2
\alpha_c \sin \left(n_a \chi t\right)$. If the input in mode $a$
is either the Fock state $|0\rangle$ or $|1\rangle$, the
respective output states of the probe mode $c$ are the coherent
states $|\alpha_c\rangle$ or $|\alpha_c e^{i \chi t}\rangle$.
Using the momentum quadrature measurement, the probability of
misidentifying one of these states for one another is then
$P_\text{error}=\frac{1}{2}\text{erfc}(\text{SNR}_Y/2\sqrt{2})$.
A signal-to-noise ratio of $\text{SNR}_Y=4.6$ would thus give
$P_\text{error} \sim 10^{-2}$. To achieve the necessary phase
shift we require $\alpha_c \sin \left(\chi t\right)\approx 2.3$,
which can be achieved in a number of ways dependent upon the range
of values available for $\alpha_c$ and $\chi t$. For example, we
could choose $\alpha_c\gg 2.3$ with $\chi t$ small and satisfy the
above inequality; alternatively we could choose $\chi t =\pi/2$
with $\alpha_c=2.3$. The particular regime chosen depends on the
strength of the Kerr nonlinearity achievable in the physical
system.
\begin{figure}[!htb]
\includegraphics[scale=0.4]{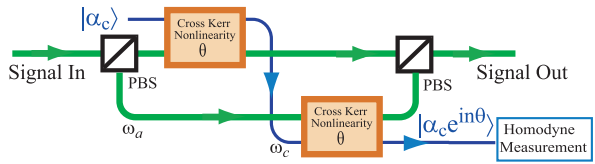}
\caption{Schematic diagram of a polarization-preserving photon
number quantum non-demolition detector based on a pair of identical
cross-Kerr optical nonlinearities. The signal mode is a Fock state
with an unknown polarization, which is resolved into orthogonal
polarization states by a polarizing beam splitter. The phase shift
applied to the probe mode is proportional to $n_a$, independent of
the polarization of the signal mode.}
\label{qnd_pp}
\end{figure}

Figure~\ref{qnd_pp} generalizes the detector shown schematically
in Fig.~\ref{qnd_ideal} to the case where the polarization of
the input state is resolved into different paths by a polarizing
beamsplitter. In general we may wish to apply different phase
shifts to the two distinct polarizations, but in the case shown in
Fig.~\ref{qnd_pp} an identical phase shift is applied to each
path, so that the detector is insensitive to the polarization of
the input state. This is a particularly useful approach when the
efficiency of the EIT system and/or the optical propagation path
(e.g., as provided by a photonic crystal waveguide optimized for
either TE or TM modes) is polarization-dependent.

\begin{figure}[!htb]
\includegraphics[scale=0.25]{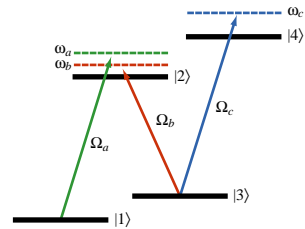}
\caption{Schematic diagram of the interaction between
a four-level $\mathcal{N}$ atom and a nearly resonant three-frequency
electromagnetic field. We note that the annihilation of a photon of
frequency $\omega_k$ is represented by the complex number $
\Omega_k$.}
\label{fourlevel}
\end{figure}

We now address the generation of the cross-Kerr nonlinearity
required to perform the QND measurement. We consider a  model
(depicted in Fig.~\ref{fourlevel}) of the nonlinear electric
dipole interaction
between three quantum electromagnetic radiation fields with
angular frequencies $\omega_a$, $\omega_b$, $\omega_c$ and a
corresponding four-level $\mathcal{N}$ atomic system\cite{beau03}.
Mode $b$ should be thought of as a pump or coupling field: by
choosing the correct conditions, we can factor both the coupling
field and the atom out of the evolution of the atom-field system,
creating an effective cross-Kerr nonlinear interaction between
modes $a$ and $c$. The effective vacuum Rabi frequency for each
mode is defined as $\left|\Omega_k\right|^2 = (\sigma_k/\eta_k
\mathcal{A})\, A_k\, \Delta \omega_k/8 \pi$, where $\sigma_k
\equiv 3 \lambda_k^2/2 \pi$ is the resonant atomic absorption
cross section at wavelength $\lambda_k \cong 2 \pi
c/\omega_{k}$\cite{cohe92}, $\eta_k$ is the refractive index of
the waveguide material, $\mathcal{A}$ is the effective laser mode
cross-sectional area, $A_k \equiv f_k e^2 \omega_k^2/2 \pi
\epsilon_0 m_e c^2$ for a transition with oscillator strength
$f_k$, and $\Delta \omega_k$ is the bandwidth of the profile
function describing the adiabatic interaction of a pulsed laser
field with a stationary atom\cite{blow90,chan02,domo02}.

It is difficult to achieve a substantial vacuum Rabi frequency
using free-space fields\cite{vane00}, but encapsulating one or
more atoms in a waveguide (such as a line defect in a photonic
crystal structure) allows field transversality to be maintained at
mode cross-sectional areas on the order of $\mathcal{A} \approx
(\lambda/3 \eta)^2$. Consider, then, a two-dimensional photonic
crystal waveguide constructed from diamond thin film ($\eta =
2.4$), with nitrogen-vacancy color centers fabricated in the
center of the waveguide channel\cite{hemm01,shah02}.  The optical
transition at 637~nm in NV-diamond has an oscillator strength of
approximately 0.12, within a factor of three of rubidium, which
has been used successfully to demonstrate cross-Kerr
nonlinearities at low light levels\cite{harr99,harris04}. An EIT
transmission window of about 8~MHz has been observed
experimentally\cite{hemm01}, so a pulse with $\Delta \omega_k/2
\pi \lesssim 5$~MHz should propagate through this window with
negligible loss. The corresponding vacuum Rabi frequency is
therefore $\Omega \approx 3.6$~MHz.

We consider a number $N$ of $\mathcal{N}$ atoms, fixed and
stationary within a cylinder that is narrow but long compared to
the optical wavelengths, with the three frequency modes of the
system driven by Fock states containing $n_a$, $n_b$, and $n_c$
photons, respectively. If the durations of the three pulse
envelope functions are long compared to the lifetime of atomic
level $| 2\rangle$, the evolution of the amplitude where all of
the atoms are in the ground state $\ket{1}$ is simply given by
$\left|1, n_a, n_b, n_c\right\rangle \longrightarrow e^{-i W t}
\left|1, n_a, n_b, n_c\right\rangle$. In general, $W$ is complex
(see Eq.\ (139) in Ref.\ \cite{beau03}); other states contribute
to the full evolution and there is photon absorption and loss in
the system. However, for our detector we require the probability
of even single photon loss from mode $a$ to be very small, with a
real $W$ preserving the norm of $\left|1, n_a, n_b,
n_c\right\rangle$. For this, we assume that the laser frequencies
$\omega_a$ and $\omega_b$ are both precisely tuned to the
corresponding atomic transition frequencies, so that the EIT Raman
resonance condition is satisfied. Then, in the weak-signal regime,
we assume that the decoherence rate $\gamma_k$ is dominated by
spontaneous emission from atomic level $|k\rangle$, and that
$|\Omega_a| \lesssim \gamma_2$, so that $W$ is given by
 \begin{equation}\label{w}
W = \frac{N\, \left|\Omega_a\right|^2 \left|\Omega_c\right|^2 n_a
n_c }{\nu_c \left|\Omega_b\right|^2 n_b + i \left( \gamma_{4}
\left|\Omega_b\right|^2 n_b + \gamma_2 \left|\Omega_c\right|^2 n_c
\right)} ,
 \end{equation}
where $\nu_c \equiv \omega_c - \omega_{43}$. In principle, in this
regime we must have $\nu_c |\Omega_b|^2 n_b \gg \gamma_4
|\Omega_b|^2 n_b + \gamma_2 |\Omega_c|^2 n_c$ to obtain a nearly
real $W$ and a low residual absorption. As we shall show below, in
practical cases where $|W t| \ll 1$, this constraint can be
substantially relaxed. For NV-diamond, $\gamma_2^{-1} = 2 \times
25$~ns \cite{beve01,beve02}, and the spin decoherence lifetime is
0.1~ms \cite{shah02,hemm01}, so for $N \lesssim 10000$, dephasing
can be neglected and $|\Omega_a|/\gamma_2 \approx 1$.

Under these conditions, the state $\left|1, n_a, n_b,
n_c\right\rangle$ simply acquires a phase-shift which is the basis
for the emergence of the approximate cross-Kerr
non-linearity\cite{beau03}. In general, when the pump and probe
fields are intense coherent states (parameterized by $\alpha_b$
and $\alpha_c$, respectively), the evolution of the state $\ket{1,
n_a, \alpha_b, \alpha_c}$ can be approximated as $\ket{1, n_a,
\alpha_b, \alpha_c e^{-i n_a (\theta - i \kappa)}}$\cite{beau03},
where the angle $\theta$ and the residual absorption $\kappa$ is
defined for the case $\left|\Omega_b\right|^2 =
\left|\Omega_c\right|^2$ by
\begin{equation}\label{thetadef}
\theta - i \kappa = \frac{N\, \left|\Omega_a\right|^2 t }{\nu_c
\left|\alpha_b\right|^2 + i \left( \gamma_{4}
\left|\alpha_b\right|^2 + \gamma_2 \left|\alpha_c\right|^2
\right)} .
 \end{equation}
This is equivalent to a damped evolution generated by the
cross-Kerr Hamiltonian of (\ref{crossKerreq}), with $\theta=\chi
t$.

What values of $\theta$---and, therefore, $\text{SNR}_Y$---are
achievable? To establish an estimate we need to make several
assumptions about the physical system and its geometry.  We assume
that the interaction region (where the light and $\mathcal{N}$
atoms interact) is encapsulated within the photonic crystal
waveguide described above, and that the pulses have weakly
super-Gaussian profiles so that the bandwidth-interaction time
product is $\Delta \omega_k t \approx 3 \pi$, giving
$\left|\Omega_a\right|^2 t \approx 81\, \eta\, \gamma_2/8 \pi$.
Suppose now that the largest phase shift that can be applied in
practice (without significantly distorting the signal pulse) is
$\theta_\text{max}$, and that the corresponding value of
$\alpha_c$ needed to obtain a given signal-to-noise ratio is
therefore $\alpha_c = \text{SNR}_Y/2\, \theta_\text{max}$ (if 
$\theta_\text{max} \ll 1$). Using
Eq.~(\ref{thetadef}), we can calculate the minimum number of atoms
and the corresponding minimum detuning $\nu_c$ needed to generate
this phase shift, by taking the real part 
and solving for $\nu_c$ explicitly. As $\nu_c$ 
is required to be real we find
\begin{eqnarray}
N_\text{min} &=& \frac{2 \theta_\text{max}}{\left|\Omega_a\right|^2 t}\,
\left( \gamma_{4} \left|\alpha_b\right|^2 + \gamma_2 \left|\alpha_c\right|^2\right) 
\nonumber \\
&=& \frac{\gamma_2 \text{SNR}_Y^2 }{2 \theta_\text{max} \left|\Omega_a\right|^2 t} 
\left(\frac{\gamma_4 \left|\alpha_b\right|^2}{\gamma_2 \left|\alpha_c\right|^2} + 1 \right)
\end{eqnarray}
with
\begin{equation}
\nu_{c\text{min}} = \frac{N_\text{min} \left|\Omega_a\right|^2 t}{2\left|\alpha_b\right|^2 \theta_\text{max}}=
\frac{\gamma_{4} \left|\alpha_b\right|^2 + \gamma_2 \left|\alpha_c\right|^2}{\left|\alpha_b\right|^2}
\end{equation}
When we choose $N = N_\text{min}$ and $\nu_c = \nu_{c\text{min}}$, we
find that $\kappa = \theta$, so that the state $\ket{1, n_a,
\alpha_b, \alpha_c}$ evolves according to $\ket{1, n_a, \alpha_b,
\alpha_c e^{-(1 + i) n_a \theta}}$. Therefore, when $\theta \ll
1$, the residual absorption of a signal photon in mode $a$ can
also be made intrinsically small, even though the detuning is not
large compared to the absorption linewidth\cite{beau04b}.

\begin{figure}[!htb]
\includegraphics[scale=0.5]{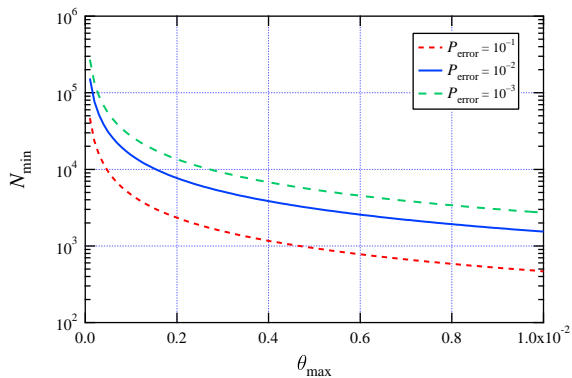}
\caption{Plot of the minimum number of NV-diamond color centers
needed to generate the phase shift $\theta_\text{max}$ for three
different values of the error probability $P_\text{error}$ in the
case $n_a = 1$. We have chosen $\langle n_b \rangle = 10 \langle
n_c \rangle$, requiring a minimum detuning $\nu_{c\text{min}}/\gamma_2 =
1.1$.} \label{qnd_min}
\end{figure}

Figure~\ref{qnd_min} shows the minimum number of NV-diamond color
centers as a function of the maximum single-photon phase angle for
three different values of the error probability $P_\text{error}$.
We note that $\gamma_4 = \gamma_2$ for the optical transitions in
NV-diamond, and for convenience we have chosen $\langle n_b
\rangle = 10 \langle n_c \rangle$, requiring a minimum detuning
$\nu_{c\text{min}}/\gamma_2 = 1.1$ in all three cases.  Note that the minimum
number of atoms needed to obtain a given phase shift decreases as
the phase shift increases, because the constraint that the
signal-to-noise ratio remain constant allows the values of
$|\alpha_b|$ and $|\alpha_c|$ to decrease as $\theta_\text{max}$
increases. As an example, we choose $P_\text{error} = 0.01$ and
$\theta_\text{max} = 0.01$ radians, requiring $\langle n_c \rangle
= 5.6 \times 10^4$ to maintain $\text{SNR}_Y = 4.6$. Therefore, $N
\approx 1600$ is sufficient to achieve the desired phase shift,
resulting in a residual absorption of less than 1\%.

There is considerable flexibility in the engineering design
parameter space for this implementation of the QND detector. For
example, if we choose $\theta_\text{max} = 0.1$, following the
design procedure outlined above for $P_\text{error} = 0.01$ leads
to a residual absorption of almost 10\% for $N \approx 160$ and
$\langle n_c \rangle \approx 560$. However, if we increase the
number of atoms to 800 and the detuning to $\nu_c = 11 \gamma_2$,
then the absorption is reduced to 1\%. Note also that the detector
can perform a QND measurement on a Fock state with $n_a > 1$ with
single-photon resolution. For example, as $n_a$ increases from 1
to 2, the phase shift $n_a \theta$ doubles, and the SNR also
increases from 4.6 to 9.2 for constant $\alpha_c$. The detector
sensitivity improves until the phase shift becomes so large that
one of two fundamental limits is reached: either the SNR decreases
below the 1\% error threshold, or the strong nonlinear interaction
begins to significantly distort the pulse profile of the signal
Fock state.

In summary, we have presented a scheme for a highly
efficient photon number quantum non-demolition detector (with
single-photon resolution) based on the cross-Kerr nonlinearity
produced by an EIT condensed matter system
with approximately 1600 color centers. We have explored several
different operating regimes, and we have examined
in detail the performance of the detector for an NV-diamond
photonic crystal waveguide system. In particular, we have shown
that efficient detection is possible with small phase shifts,
which will likely be necessary to ensure that the EIT optical
nonlinearity doesn't distort the pulse envelope of the signal
state. Future modelling will address detector performance
for pulsed Fock state profile functions, and much
experimental work remains to be done to implement such a detector.
For example, fabricating EIT atomic or molecular systems into a
dielectric waveguide is challenging but feasible. A method for
orienting the color center spins uniformly in such a
condensed-matter system must be found, and spatial hole-burning
techniques will be needed to overcome the effects of inhomogeneous
broadening on the transparency of a condensed matter
medium\cite{turu02}. The overall efficiency of the detector is
likely to be limited by the efficiency of the homodyne measurement
of the phase shift, which will depend on the degree to which the
homodyne detector can be spatiotemporally mode-matched to a
single-photon signal. Nevertheless, EIT provides us with the best
known candidate mechanism for the implementation of the original
QND proposal by Imoto, Haus, and Yamamoto\cite{imot85}, and even a
weak nonlinearity could allow efficient reuse of resources in
linear optics quantum computation schemes.

\noindent {\em Acknowledgments}: This work was supported by the
European Project RAMBOQ. KN acknowledges support in part from
MPHPT, JSPS, Asahi-Glass and Japanese Research Foundation for
Opto-Science and Technology research grants. WJM acknowledges support 
in the form of a JSPS fellowship.

\end{document}